\documentclass[pra,twocolumn,showpacs]{revtex4}   
\def\ausilio{}

\font\titolo=cmbx12 2%
\font\sc=cmcsc10\font\css=cmcsc8%
\font\ottorm=cmr8%
\def\st{\scriptstyle}%
\font\tenmib=cmmib10 \font\eightmib=cmmib8
\font\sevenmib=cmmib7\font\fivemib=cmmib5 
\font\ottoit=cmti8\font\fiveit=cmti5\font\sixit=cmti6
\font\fivei=cmmi5\font\sixi=cmmi6\font\ottoi=cmmi8
\font\ottorm=cmr8
\font\ottosy=cmsy8\font\sixsy=cmsy6\font\fivesy=cmsy5
\font\ottobf=cmbx8\font\sixbf=cmbx6\font\fivebf=cmbx5%
\font\ottocss=cmcsc8%

\def\ottopunti{\def\rm{\fam0\ottorm}\def\it{\fam6\ottoit}%
\def\bf{\fam7\ottobf}%
\textfont1=\ottoi\scriptfont1=\sixi\scriptscriptfont1=\fivei%
\textfont2=\ottosy\scriptfont2=\sixsy\scriptscriptfont2=\fivesy%
\textfont4=\ottocss\scriptfont4=\sc\scriptscriptfont4=\sc%
\textfont5=\eightmib\scriptfont5=\sevenmib\scriptscriptfont5=\fivemib%
\textfont6=\ottoit\scriptfont6=\sixit\scriptscriptfont6=\fiveit%
\textfont7=\ottobf\scriptfont7=\sixbf\scriptscriptfont7=\fivebf%
\setbox\strutbox=\hbox{\vrule height7pt depth2pt width0pt}%
\normalbaselineskip=9pt\rm}
\let\nota=\ottopunti%

\mathchardef\Ba   = "050B  
\mathchardef\Bb   = "050C  
\mathchardef\Bg   = "050D  
\mathchardef\Bd   = "050E  
\mathchardef\Be   = "0522  
\mathchardef\Bee  = "050F  
\mathchardef\Bz   = "0510  
\mathchardef\Bh   = "0511  
\mathchardef\Bthh = "0512  
\mathchardef\Bth  = "0523  
\mathchardef\Bi   = "0513  
\mathchardef\Bk   = "0514  
\mathchardef\Bl   = "0515  
\mathchardef\Bm   = "0516  
\mathchardef\Bn   = "0517  
\mathchardef\Bx   = "0518  
\mathchardef\Bom  = "0530  
\mathchardef\Bp   = "0519  
\mathchardef\Br   = "0525  
\mathchardef\Bro  = "051A  
\mathchardef\Bs   = "051B  
\mathchardef\Bsi  = "0526  
\mathchardef\Bt   = "051C  
\mathchardef\Bu   = "051D  
\mathchardef\Bf   = "0527  
\mathchardef\Bff  = "051E  
\mathchardef\Bch  = "051F  
\mathchardef\Bps  = "0520  
\mathchardef\Bo   = "0521  
\mathchardef\Bome = "0524  
\mathchardef\BG   = "0500  
\mathchardef\BD   = "0501  
\mathchardef\BTh  = "0502  
\mathchardef\BL   = "0503  
\mathchardef\BX   = "0504  
\mathchardef\BP   = "0505  
\mathchardef\BS   = "0506  
\mathchardef\BU   = "0507  
\mathchardef\BF   = "0508  
\mathchardef\BPs  = "0509  
\mathchardef\BO   = "050A  
\mathchardef\BDpr = "0540  
\mathchardef\Bstl = "053F  

\global\newcount\numsec\global\newcount\numapp
\global\newcount\numfor\global\newcount\numfig
\global\newcount\numsub
\def\veroparagrafo{\number\numsec}\def\veraformula{\number\numfor}
\def\veraappendice{\number\numapp}\def\verasub{\number\numsub}
\def\verafigura{\number\numfig}

\def\senondefinito#1{\expandafter\ifx\csname#1\endcsname\relax}

\def\SIA #1,#2,#3 {\senondefinito{#1#2}%
\expandafter\xdef\csname #1#2\endcsname{#3}\else
\write16{???? ma #1#2 e' gia' stato definito !!!!} \fi}

\def \Fe(#1)#2{\SIA fe,#1,#2 }
\def \Fp(#1)#2{\SIA fp,#1,#2 }
\def \Fg(#1)#2{\SIA fg,#1,#2 }

\def\Section(#1,#2){\advance\numsec by 1\numfor=1\numsub=1\numfig=1%
\SIA p,#1,{\veroparagrafo} %
\hbox to \hsize{\titolo\hfill \number\numsec. #2\hfill%
\expandafter{\alato(sec. #1)}}\*}

\def\Appendix(#1,#2){\advance\numapp by 1\numfor=1\numsub=1\numfig=1%
\SIA p,#1,{A\veraappendice} %
\hbox to \hsize{\titolo Appendix A\number\numapp. #2\hfill%
\expandafter{\alato(app. #1)}}%
\*%
}

\def\etichetta(#1){(\veroparagrafo.\veraformula)%
\SIA e,#1,(\veroparagrafo.\veraformula) %
\global\advance\numfor by 1%
}

\def\etichettaa(#1){(A\veraappendice.\veraformula)%
\SIA e,#1,(A\veraappendice.\veraformula) %
\global\advance\numfor by 1%
}

\def\getichetta(#1){\veroparagrafo.\verafigura%
\SIA g,#1,{\veroparagrafo.\verafigura} %
\global\advance\numfig by 1%
}

\def\etichettap(#1){\veroparagrafo.\verasub%
\SIA p,#1,{\veroparagrafo.\verasub} %
\global\advance\numsub by 1%
}

\def\Eq(#1){\eqno{\etichetta(#1)\alato(#1)}}
\def\eq(#1){\etichetta(#1)\alato(#1)}
\def\Eqa(#1){\eqno{\etichettaa(#1)\alato(#1)}}
\def\eqa(#1){\etichettaa(#1)\alato(#1)}
\def\eqg(#1){\getichetta(#1)\alato(fig. #1)}
\def\sub(#1){\0\palato(p. #1){\bf \etichettap(#1).}}
\def\asub(#1){\0\palato(p. #1){\bf \etichettapa(#1).}}
\def\apprif(#1){\senondefinito{e#1}%
\eqv(#1)\else\csname e#1\endcsname\fi}

\def\equv(#1){\senondefinito{fe#1}$\clubsuit$#1%
\write16{eq. #1 non e' (ancora) definita}%
\else\csname fe#1\endcsname\fi}
\def\grafv(#1){\senondefinito{fg#1}$\clubsuit$#1%
\write16{fig. #1 non e' (ancora) definito}%
\else\csname fg#1\endcsname\fi}
\def\secv(#1){\senondefinito{fp#1}$\clubsuit$#1%
\write16{par. #1 non e' (ancora) definito}%
\else\csname fp#1\endcsname\fi}

\def\eqo{{\global\advance\numfor by 1}}
\def\equ(#1){\senondefinito{e#1}\equv(#1)\else\csname e#1\endcsname\fi}
\def\graf(#1){\senondefinito{g#1}\grafv(#1)\else\csname g#1\endcsname\fi}
\def\figura(#1){{\css Figura} \getichetta(#1)}
\def\secc(#1){\senondefinito{p#1}\secv(#1)\else\csname p#1\endcsname\fi}
\def\sec(#1){{\secc(#1)}}
\def\refe(#1){{[\secc(#1)]}}

\def\BOZZA{
\def\alato(##1){\rlap{\kern-\hsize\kern-.5truecm{$\scriptstyle##1$}}}
\def\palato(##1){\rlap{\kern-.5truecm{$\scriptstyle##1$}}}
}

\def\alato(#1){}
\def\galato(#1){}
\def\palato(#1){}

{\count255=\time\divide\count255 by 60 \xdef\hourmin{\number\count255}
        \multiply\count255 by-60\advance\count255 by\time
   \xdef\hourmin{\hourmin:\ifnum\count255<10 0\fi\the\count255}}

\def\oramin{\hourmin }

\def\data{\number\day/\ifcase\month\or gennaio \or febbraio \or marzo \or
aprile \or maggio \or giugno \or luglio \or agosto \or settembre
\or ottobre \or novembre \or dicembre \fi/\number\year;\ \oramin}
\setbox200\hbox{$\scriptscriptstyle \data $}

\let\a=\alpha \let\b=\beta  \let\g=\gamma    \let\e=\varepsilon
\let\z=\zeta        
\let\m=\mu                  
\let\s=\sigma \let\t=\tau   \let\f=\varphi 
      
\let\G=\Gamma \let\D=\Delta   
\let\P=\Pi

\def\CC{{\cal C}}

\let\ig=\int
\let\io=\infty
\let\==\equiv
\let\0=\noindent
\let\Dpr=\BDpr

\def\\{\hfill\break}
\def\lis#1{\overline#1}

\def\*{\vskip3mm} 
\def\ie{{\it i.e. }}
\def\eg{{\it e.g. }}

\let\dpr=\partial
 
\def\defi{\,{\buildrel def\over=}\,}

\def\V#1{{\bf#1}}
\def\media#1{{\langle#1\rangle}}
\def\fra#1#2{{#1\over#2}}
\def\crcl{\,\raise.5mm\hbox{$\st\rm o$}\,}%
\def\otto{\,{\kern-1.truept\leftarrow\kern-5.truept\to\kern-1.truept}\,}
\def\tende#1{\,\vtop{\ialign{##\crcr\rightarrowfill\crcr
 \noalign{\kern-1pt\nointerlineskip} \hskip3.pt${\scriptstyle
 #1}$\hskip3.pt\crcr}}\,}
\newdimen\xshift \newdimen\xwidth \newdimen\yshift \newdimen\ywidth

\def\ins#1#2#3{\vbox to0pt{\kern-#2\hbox{\kern#1 #3}\vss}\nointerlineskip}

\def\eqfig#1#2#3#4#5{
\par\xwidth=#1 \xshift=\hsize \advance\xshift
by-\xwidth \divide\xshift by 2
\yshift=#2 \divide\yshift by 2%
{\hglue\xshift \vbox to #2{\vfil
#3 \includegraphics{#4.eps}
}\hfill\raise\yshift\hbox{#5}}}

\def\8{\write12}

\input #1.sty \closein13\fi
\input \jobname.sty \closein14 \fi

\newcommand\revtex{{R\kern-0.4mm\lower0.5mm\hbox{E}\kern-0.4mm V\kern-0.3mm%
\lower0.5mm\hbox{T}\kern-0.4mm E\kern-.3mm \lower0.5mm\hbox{X}}}

\usepackage{eqalignno}

\ausilio

\begin{document}
\title{Fluctuation relation, fluctuation theorem, thermostats and
  entropy creation in non equilibrium statistical
Physics}
\*

\author{Giovanni Gallavotti}
\affiliation{I.N.F.N. and Fisica, Universit\`a di Roma 1}

\date{\today}

\begin{abstract} A unified viewpoint is presented in margin to the
{\rm ``Work, dissipation and fluctuations in nonequilibrium physics''}
Bruxelles 22-25 March, 2006, where the topics were discussed by
various authors and it became clear the need that the very different
viewpoints be consistently presented by their proponents
\end{abstract}

\maketitle

\0{\bf 1. \it Chaotic dynamics}
\numsec=1\numfor=1\*

The {\it fluctuation relation} is a general symmetry property of
mechanical systems which should hold under the only assumption that
the system motions are {\it chaotic}: it reflects the {\it time
reversal symmetry}.

Time reversal symmetry means a smooth isometry $I$ of phase space
which anticommutes with the time evolution map $S_t$: \ie
$IS_t=S_{-t}I$. Therefore the familiar operation of time reversal $T$
or, even better, $TCP$ would be always valid in fundamental models,
\cite{Ga98}, and therefore the time reversal symmetry can be an issue
only if one deals with phenomenological models in which dissipation is
empirically introduced: as in the case of Navier-Stokes equations for
fluids, for instance.

The discovery, published in the paper \cite{ECM93}, of the fluctuation
relation has led to renewed efforts towards the formulation of a
theory of nonequilibrium stationary states. In the paper a link is
attempted with the earlier proposal \cite{Ru80} for the description of
the probability distribution for chaotic stationary states in
fluids. Although the paper was a real breakthrough, the original
argument needed to be made precise. A direct connection with
\cite{Ru80} was established and called ``{\it fluctuation theorem}''
in \cite{GC95}: where it was shown how the paradigm of chaotic evolution
constituted by the hyperbolic (also called ``Anosov'') systems allowed
for a precise formulation of sufficient conditions under which the
fluctuation relation held.

The latter proof has been considered interesting because in a sense
the hyperbolic evolutions perform for chaotic systems the role plaid
by the harmonic oscillators for ordered systems. From these works
emerged the interest for Physics of two new fundamental concepts, the
{\it chaotic hypothesis} and a general mechanical notion for the {\it
entropy creation}.

The first is an extension of the proposals, \cite{Ru80}, that
identified the probability distributions forming the ensembles
suitable to give the statistical properties of states of turbulent
fluids with the special class of distributions, well known and studied
in the theory of dynamical systems, called SRB distributions.

The extension can be formulated, \cite{Ru80,GC95}, in the form a
hypothesis: \*

\0{\bf Chaotic Hypothesis}: {\it Motion on the attracting set of a
  chaotic system can be regarded as ``hyperbolic''.}  \*

This hypothesis is very ambitious as it should be seen as an extension
of the {\it ergodic hypothesis}, which it implies if applied to a
system which is isolated and subject only to conservative forces. 

It allows us to state existence of time averages of mechanical
observables $x\to F(x)$, identified with functions of the phase space
point $x$ representing the microscopic state of the system, \ie the
existence of the limits $\lim_{T\to\io} \fra1T \sum_{j=0}^{T-1}
F(S^jx)$ for all initial data $x$ chosen in the vicinity of the
attracting set, setting aside a set of $0$ phase space volume. It also
implies that, the limit is {\it independent} of $x$ (apart from the
zero volume possible exceptions) and therefore define a {\it
statistics}, \ie a probability distribution $\m$ such that

$$\lim_{T\to\io} \fra1T \sum_{j=0}^{T-1} F(S^jx)=\ig\m(dy) \,F(y)\Eq(e1.1)$$
In other words for chaotic systems motions have a well defined
statistics, \ie a probability distribution that allows us to define
(in principle) the time averages of the observables.

The nontriviality of the above statements becomes perhaps more clear
in the case of systems subject to steady dissipation. Their stationary
states cannot be described by statistics which are not {\it singular}:
\ie which attribute probability $1$ to sets of $0$ phase space
volume. 

This is a seemingly odd situation: we are interested in data randomly
chosen with a probability distribution with density on phase space
and, yet, they evolve with a statistics which is singular. Such a
situation, however, appears to be quite clearly correct as soon as
simulations are attempted in {\it virtually any} system which exhibits
chaos (\ie positive Lyapunov exponents) and is subject to some kind of
dissipation: hence it is simply accounted for by the chaotic
hypothesis, see for instance \cite{BGG97}.

However technically the hypothesis is far more rich of
implications. In fact the distribution $\m$ is, in hyperbolic systems,
identified with the SRB distribution, which is not well known but it
should be viewed as a generalization of the microcanonical
distribution for isolated systems. In a sense that can be made very
precise it is the distribution that gives equal probability to cells
into which the phase space can be imagined discretized,
\cite{Ga95a,Ga95c,Ga99}. It allows to give a mathematical formula for
the averages in \equ(e1.1) and even to give a precise definition of
{\it coarse grained cells} in phase space, \cite{Ga95a,Ga01,Ga06b}.

Without entering here in more technical details this means that we
have, from the theory of the SRB distributions for hyperbolic systems,
expressions for the averages in terms of mechanical quantities. The
latter can therefore be used to derive general relations between
observables averages {\it even though, as it is virtually always the
case, we cannot hope to compute their actual vlaues}.

Just as in equilibrium where we can write the averages as integrals
with respect to the Liouville distribution on the energy surface (or
the canonical one) but we can hardly compute them: nevertheless we can
establish, by using the formal expressions, general relations which
turn out to be extremely interesting precisely because of their
generality.  A celebrated example is Boltzmann's {\it heat theorem},
\ie the second law of equilibrium thermodynamics, as a consequence of
the assumption that the statistics of motion (of an isolated system
subject to conservative forces) is the microcanonical distribution.

The chaotic hypothesis has a rather general consequence which should
be seen as a generalization, at any distance from equilibrium, of
Onsager-Machlup fluctuations theory near equilibrium,
\cite{OM53a,OM53b,Ga96a}.

Consider a mechanical system of particles described by a generic
equation of motion for the representative point $x$ in phase space

$$\dot x= f(x)\Eq(e1.2)$$
where $x$ denotes the position and velocity components. The equation
will be a model describing a finite system on which external non
conservative forces act.

Therefore the equation will be non
Hamiltonian and phase space volume will not be conserved. As a
consequence the {\it divergence} 

$$\s(x)=-\sum_i\dpr_{x_i} f_i(x)\Eq(e1.3)$$
{\it does not vanish}. However in presence of time reversal symmetry it
will be odd under time reversal $\s(x)=-\s(Ix)$. The system will be
called {\it dissipative} if even the time average $\s_+=\lim_{T\to\io}
\fra1T\ig_0^T \s(S_tx)dt=\ig \m(dy)\s(y)$ of $\s$ is {\it positive}.

In a dissipative system, $\s_+>0$, in the stationary state described
by the statistics $\m$, consider the probability that $f_j(S_t
x)\sim\f(t)$ for $t\in [-\fra12\tau,\fra12\tau]$ where $t\to\f(t)$ is
a prescribed {\it pattern}.  The symbol $\sim$ means that
$|f(S_tx)-\f(t)|<\e$ for some very small $\e$ (see \cite{Ga97,Ga00,
Ga02} for a quantitative form of the notion of ``very small''). Define
the {\it dimensionless phase space contraction} $\s(x)$ as the
divergence of the equations of motion changed in sign.  Suppose that
the average (with respect to the statistics $\m$, \ie the integral of
$\s(x)$ with respect to $\m$, see \equ(e1.1)) phase space contraction
$\s_+$ is positive $\s_+>0$ and define
$p=\fra1T\ig_{-\fra\t2}^{\fra\t2} \fra{\s(S_t x)}{\s_+}dt$.  Suppose
that $f_1(x),\ldots,f_n(x)$ are $n$ observables with defined parity
under time reversal, suppose odd for definiteness: $f_i(Ix)=-
f_i(x)$. Then \*

\0{\bf Fluctuation relation} {\it Suppose that the $n$ observables
form a complete set in the sense that the average phase space
contraction $p$ is determined by the patterns followed by the
observables $f_1,\ldots, f_n$ (for instance $\s$ is one among the
$f_j$ or $p$ is the sum of the averages of some of the $f_j$). Then if
$\s_+>0$ and the time evolution is reversible there exists $p^*\ge1$ and

$$\eqalignno{&
\fra{P_\t(\hbox{\rm for all}\ j, {\rm and}\,
t\in[-\fra12\t,\fra12\t]
\,: f_j(S_tx)\sim\f_j(t))}
{P_\t(\hbox{\rm for all}\ \,j, {\rm and}\,
t\in[-\fra12\t,\fra12\t]\,: f_j(S_tx)\sim -\f_j(-t))}
=\cr&\kern1cm
=e^{\,p\,\s_+\, \t+O(1)},\qquad |p|<p^*
&\rm\eq(e1.4)\cr}
$$
where $P_\t(E)$ denotes the probability of the event $E$  with respect
to the statistics SRB of the motions.
}
\*

\0{\it Remarks:} 
(1) In particular this holds for the single observable $\s(x)/\s_+$,
\cite{GC95, Ge98}, and {\it it is a theorem for hyperbolic systems}. This is the
form in which the fluctuation relation was discovered, \cite{ECM93}, and the
above is an {\it extension} of it under the same assumptions, \cite{Ga97,Ga00}.
\\
(2) The extension was found first in the special case $f_1=\s(x)/\s_+,
    f_2=\dpr_E \s(x)$ where it has been shown to imply the Onsager
    reciprocity and Green-Kubo formulae, \cite{Ga96a}.
\\
(3) Another particular case is obtained by considering the probability
    that the averages $\lis f_1,\ldots, \lis f_n$, over the time
    interval $(-\fra\t2,\fra\t2)$ of the considered observables,
    have a given value $a_1,\ldots, a_n$ with $p$ determined by
    $a_1,\ldots, a_n$. Then for $|p|<p^*$

$$\fra{P( \lis f_1\sim a_1,\ldots, \lis f_n\sim a_n)}{ P( \lis f_1\sim
-a_1,\ldots, \lis f_n\sim -a_n)}=e^{p\s_+ \t+O(1)}\Eq(e1.5)$$
which is a very surprising relation because of the arbitrariness of
the observables $f_j$ {\it which do not appear in the r.h.s.} except
through their function $p$. 
The above relation appeared recently in the
context of Kraichnan's theory of passive scalars in a case in which
$p=\sum_i a_i$, \cite{CDG06}.
\\
(4) A mahematically precise form of the theorem, \cite{GC95,Ga95b}, is to say
    that for $|p|<p^*$ the probability that $p\in\D$ has the form
    $\exp(\t\max_{p\in\D}\z(p)+O(1))$ and the function $\z(p)$, which
    for hyperbolic systems is known to be analytic and convex in a
    natural interval of
    definition $(-p^*,p^*)$ (and $-\io$ outside it) satisfies, for
    $|p|<p^*$ the symmetry property

$$\z(-p)=\z(p)-p\s_+\Eq(e1.6)$$
\\
(5) A further consequence is that the stationary average of $\exp
    \ig_0^\t \s(S_tx)dt$ satisfies

$$\media{e^{\ig_0^\t \s(S_tx)dt}}\sim1\Eq(e1.7)$$
where $\sim1$ means that the quantity is {\it bounded} as $\t\to\io$
({\it Bonetto}'s formula, \cite[Eq.(9.10.4)]{Ga00}): see below 
for a hint to possible applications and for its similarity to
Jarzynski's formula, \cite{Ja97,Ja99}.
\\
(6) The chaotic hypothesis and the fluctuation relation have been
tested quite extensively, starting with \cite{BGG97}, and it has almost
become a test of the correctness of the computer programs simulations
of chaotic systems rather than
a formula to be tested. The situation is quite different with
experiments where a lot of difficulties arise, on a case by case
basis, in setting up experiments and interpreting them. Nevertheless
there have been several attempts and it can be hoped that more will
come, \cite{CV03a,FM04,VCC04,GC05,CHGLPR03,BCG06}.
\\
(7) The formulae above hold for time evolutions described by maps
iterations as well as for those described by differential equations.
It is worth stressing that most works, in particular \cite{GC95}, deal
with discrete time evolutions. The case of continuous time is
considered less frequently, and for the first time it has been
formulated as a theorem in \cite{Ge98}. The case of maps is possibly
closer to applications as observations are usually done when some
timing events occurr and evolution appears as a map between timing
events. However a close examination of the relations between the
continuous and discrete cases reveals a number of delicate properties
(particularly in the case in which singular forces may be acting, like
Lennard-Jones type of repulsive cores) which if neglected may lead to
errors, as exemplified in \cite{BGGZ05}.
\*

\0{\bf2.} {\it Entropy creation}
\numsec=2\numfor=1\*

 The result in the previous section hints at another major point of
 the research in the last 20 years of the '900's: in the '980's 
 a concrete model of a thermostat became necessary to perform
 simulations of molecular dynamics in systems out of equilibrium.

The ``Nos\'e--Hoover'', the ``isokinetic'' or the ``isoenergetic''
thermostats are prominent examples that were put forward and employed
to study a large number of problems: see \cite{EM90} for a thorough
discussion of the related problems. One of the results was the
discovery of the flutuation relation.

Another important byproduct was the identification of the (not yet
defined at the time) {\it entropy creation rate} with the phase space
contraction. Although the original authors quite clearly attributed to
the words they employed a meaning close to the physical one suggested
by the given names the general attitude was, it seems, to regard the
thermostat models as unphysical and, consequently, to attribute little
value to the concept of entropy creation as related in some way to the
thermodynamic entropy.

The above (extension) of the fluctuation theorem and relation,
\equ(e1.4), suggests that one should give a {\it fundamental physical 
sense to the phase space contraction}, at least in finite, time
reversible systems. Since as said above ultimately time reversal (or
the equivalent, for our purposes, TCP symmetry) is a law of nature all
models should either display the symmetry or be equivalent to
symmetric models. 

In my view this identification between phase space contraction and
entropy creation rate is an important new development,
\cite{EM90,Ru97,Ru99}, that is still not fully appreciated as it
should.

It has to be stressed that, although since more than a century we are
familiar with the entropy of equilibria and its mechanical
interpretation, no mechanical definition of entropy creation rate in a
process out of equilibrium has been proposed (or, better, accepted).
The above extended fluctuation relation is clearly saying that the
independence of the ratio in \equ(e1.5) means that the conditional
probability that a pattern occurs in presence of an average
(dimensionless) phase space contraction $p$ is the same as that of the
reverse pattern in presence of the opposite average phase space
contraction (\ie $-p$).

In other words if the entropy creation rate is reversed during a time
interval then evolution of the {\it other} observables ``proceeds
backwards'' with the same likelyhood it had to ``proceed forward''
when the average entropy production was $p$. {\it All that has to be
done to reverse the time arrow is the reverse the entropy creation
rate}.

All this leads to say that the identification of entropy creation rate
and phase space contraction has to be taken seriously. Its identity
with what one would naturally call entropy creation in the many
particular cases studied in the works summarized in \cite{EM90} and
continuing since should not be considered a curious coincidence but as
a new insight into the foundations of nonequilibrium statistical
mechanics.

Note that we are saying that in nonequilibrium entropy creation is
defined and identified with mechanical quantities: but entropy itself
is not defined (yet?), not even in stationary states. It might even
be not needed and not unambiguous, \cite{Ga00,Ga01}.

To illustrate further the new identification of entropy creation rate
with the mechanical quantity expressed by the divergence of the
equations of motion, \equ(e1.3), and its physical interest I discuss a
class of examples which, it seems to me, fully justify the mentioned
identification, \cite{Ga96,Ga06}.

Consider a mechanical system $\CC_0$ in contact with the mechanical
systems $\CC_i$, $i=1,2,\ldots$. Microscopically the state is is
described by the positions ${\bf X}_0,{\bf X}_1,\ldots,{\bf X}_n$ of the
$N_1,N_1,\ldots,N_n$ constituent particles.  The systems interact via
short range pair forces with potential energies $U_a({\bf X}_i)$,
$U_a({\bf X_i},{\bf X_0})$. There is no direct interaction between the
particles in $\CC_a,\,a>0$. 

The systems in $\CC_a,\, a>0$ should be thought as ``thermostats'' acting
on the system $\CC_0$ across the separating walls via their
mutual pair interactions.

\eqfig{210pt}{90pt}{
\ins{80pt}{60pt}{$\V X_0,\V X_1,\ldots,\V
X_n$}
\ins{43pt}{27pt}{$\st\ddot{\V X}_{0i}=-\dpr_i U_0(\V X_0)-\sum_{a}
\dpr_i U_a(\V X_0,\V X_i)+\V F_i$}
\ins{43pt}{10pt}{$\st\ddot{\V X}_{ai}=-\dpr_i U_a(\V X_a)-
\dpr_i U_a(\V X_0,\V X_i)-\a_a \dot{\V X}_{ai}$}
}{fig}{}

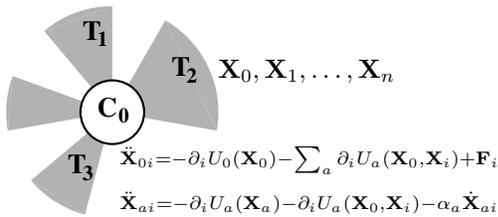
\0{Fig.1: \nota Schematic illustration of the geometry. The equations
of motion are written here assuming unit mass for the
particles.} 
\*

The thermostats {\it temperature} $T_a$ is defined to be proportional
to the kinetic energy via the Boltzmann's constant $k_B$. Setting
$K_a\=\fra12\sum_a\dot{\V X}_{a}^2\defi \fra32 k_B T_a\,N_a$, it is
supposed constant and kept such by the action of suitable
(phenomenological) forces on the $i$-th particle in $\CC_a$ of the
form $-\a_a \dot{\bf X}_{ai}$. The $\a_a$ can be taken

$$\a_a = \fra{W_a-\dot U_a}{3 N_a k_B T_a}\Eq(e2.1)$$
where where the work performed by the system on the thermostats
particles $-\sum \dot{\V X}_a\cdot\Dpr_{\V X_a}U_a(\V X_0,\V X_a)$ can
be called $W_a=\dot Q_a$ = heat given to the thermostat $\CC_a$ by the
system in $\CC_0$. The external forces $\V F_i$ are assumed to be
purely positional.

Given the above dynamical model (for heat transport) remark that it is
{\it reversible} and time reversal is just the usual velocity reversal
(because the thermostat forces are {\it even} under global velocities
change).  Furthermore the divergence of the total phase space volume
can be immediately computed and turns out to be $\s(\V X)=\s_0(\V
X)+\dot U(\V X)$ with

$$\s_0(\V X)=\sum_{a=1}^n \fra{\dot Q_a}{k_B T_a}\,\fra{3N_a-1}{3 N_a}=
\sum_{a=1}^n \fra{\dot Q_a}{k_B T_a}\Eq(e2.2)$$
where $U(\V X)=\sum_{a=1}^n \fra{\dot U_a}{k_B T_a}\,\fra{3N_a-1}{3
N_a}$, and $O(N_a^{-1})$ has been neglected in the last equality in
\equ(e2.2).

When computing time averages the ``extra term'' $\dot U$ will not
contribute because being a time derivative its average will be
$\fra1T(U(S_T\V X)-U(\V X))$ and therefore will give a vanishing
contribution for large $T$ and the average of $\sum_{a=1}^n \fra{\dot
Q_a}{k_B T_a}=\s_0$ will be the average of $\s$. If the interaction
$U$ is bounded also the fluctuations of the averages of $\s$ and $\s_0$
will coincide. In the cases in which the interactions are not bounded
(\eg Lennard-Jones repulsive cores) care has to be exercised in the
fluctuations analysis: the picture does not change except in a rather
well understood, trivial, way and this will not be discussed here,
\cite{CV03,BGGZ05}.

Note that $\s_0$ is a ``boundary term'', in the sense that it depends
on the forces through the boundaries and the forces are supposed short
range. Thus the question arises whether such kind of thermostats and
short range interactions can lead to stationary states: this is not
obvious but the ``efficiency'' of such thermostats has been
investigated in molecular dynamics simulations, \cite{AES01,GG06},
leading (not surpringly) to the result that the thermostat mechanism
in Fig.1 can lead to stationary states (even in presence of additional
positional forces stirring the particles in $\CC_0$).

Since the thermotats are regarded in equilibrium the above expression
shows that $\s(\V X)$ can be ``legitimately'' called the entropy
increase of the reservoirs: so the mechanical notion of phase space
contraction acquires a clear physical meaning: and this is a no small
achievement of a long series of works based on simulations of
molecular dynamics, \cite{EM90,Ru99}.

\*
\0{\bf 3.} {\it Comments}
\numsec=3\numfor=1\*

\0(1) Other studies of fluctuations have been proposed: they are
rather different and apply to systems which are not stationary. The
object of study are initial data {\it sampled within an
equilibrium distribution} of a Hamiltonian system and subsequently
evolved with the equations of motion of a dissipative time reversible
system. 

Then the phase space contraction averaged over a time $\t$,
$a\defi\fra1\t\ig_0^\t\s(S_t x)dt$, will be such that the probability
$P_0(a)$ {\it with respect to the initial equilibrium distribution}
for $a$ to have a given value is such that

$$\fra{P_0(a)}{P_0(-a)}=e^{a \t}.\Eq(e3.1)$$
This is an exact identity, immediately following from the
definitions. It involves {\it no error terms}, unlike the ``similar''
\equ(e1.6) that can be written also as
$\fra{P(p)}{P(-p)}=e^{p\s_+\t+O(1)}$, with $P$ the {\it probability
with respect to the stationary distribution}, which is {\it singular}
with respect to the equilibrium distributions if $\s_+>0$.

It has been claimed that, being valid for all times, it implies the
fluctuation relation, \equ(e1.6), for stationary states (at least when
the stationary state exists). This would imply a simple, direct and
{\it assumptionless} derivation of the fluctuation theorem in \equ(e1.7) and
should hold in spite of the fact that in \cite{GC95} an assumption
about the chaotic nature of the motions is needed to derive it,
together with a rather detailed understanding of the nature of chaotic
systems.

However a derivation of the fluctuation theorem \equ(e1.6) from
\equ(e3.1) involves considering \equ(e3.1) {\it after} the limit
$\t\to\io$ has been performed: a rather unclear procedure (note that
the {\it r.h.s.} depends on $\t$). Leaving aside the logical consitency
problems it should be kept in mind that in the stationary state, at
least in the interesting cases in which $\s_+>0$ and there is
dissipation, the statistics of motion will be controlled by a
distribution that has nothing to do with the initial equilibrium
distribution in which the averages in \equ(e3.1) are considered.

Therefore the claim is incorrect and it is no surprise that some kind
of chaos has to be present to obtain the fluctuation relation
\equ(e1.6). In fact one can give examples of simple systems in which
\equ(e3.1) holds for all times, the system evolves towards a
stationary state and nevertheless the \equ(e1.6) {\it does not hold},
\cite{CG99}. The confusion has crept into the literature and even
affected experiments: this can only be explained by a certain lack of
attention to the literature due to the urge to find an easy way of
testing the large fluctuations in real systems (fluids, granular
materials, or even biological systems).

Another aspect of \equ(e3.1) is that it involves $a$ rather than
$p=a/\s_+$. This is clearly a matter of convention: however care has
to be exercised because the fluctuation relation \equ(e1.6) is valid
for $|p|<p^*$ with $p^*\ge1$ being a physically nontrivial quantity,
\cite{Ga95b}. In terms of $a$ this means that it is valid for
$|a|<\s_+$. Overlooking this fact has led to think that it should hold
for all $a$'s and has led to errors in the literature. The errors are
particularly noticeable in cases in which $\s_+$ is close to zero, when
the interpretation of simulations becomes quite difficult because long
time scales become relevant. It has to be noted that $\s_+^{-1}$ is a
time scale diverging as $\s_+\to0$, see the discussion in
\cite{BGGZ05}.  \*

\0(2) A different, interesting, fluctuation result is {\it Jarzynski's
formula} which provides the means of computing the free energy
difference between two {\it equilibrium states} at the same
temperature.

Imagine to extract samples $x=(p,q)$ in phase space with a canonical
probability distribution $\m_0(dpdq)= Z_0^{-1}e^{-\b H_0(p,q)}dpdq$,
with $Z_0$ being the canonical partition function, and let
$S_{0,t}(p,q)$ be the solution of the Hamiltonian {\it time dependent}
equations $\dot p=-\dpr_q H(p,q,t),\dot q=\dpr_p H(p,q,t)$ for $0\le
t\le1$. Let $H_1(p,q)\defi H(p,q,1)$, then, \cite{Ja97,Ja99}, \*

\0{\it Consider the ``time $1$ map'' $(p',q')\defi S_{0,1}(p,q)$ and
  call $W(p',q')\defi$ $ H_1(p',q')-H_0(p,q)$ the corresponding
  variation of the energy function. Then the distribution $Z_1^{-1}
  e^{-\b H_1(p',q')}dp'dq'$ is exactly equal to the distribution
  $\fra{Z_0}{Z_1} e^{-\b W(p',q')}\m_0(dp dq)$. Hence

$$\media{e^{-\b W}}_{\m_0}=\fra{Z_1}{Z_0}=e^{-\b \D F(\b)}\Eq(e3.2)$$ 
where the average is with respect to the Gibbs distribution $\m_0$ and
$\D F$ is the free energy variation between the equilibrium states
with Hamiltonians $H_1$ and $H_0$ respectively.}
\*

\0{\it Remark:} (i) The reader will recognize in this {\it exact
identity} an instance of the Monte Carlo method. Its interest lies in
the fact that it can be implemented {\it without actually knowing}
neither $H_0$ nor $H_1$ nor the {\it protocol} $H(p,q,t)$. To evaluate
the difference in free energy bewteen two equilibrium states {\it at
the same temperature} of a system that one can construct in a
laboratory, when the system changes its energy function from $H_0$ to
$H_1$ (not necessarily explicitly known), then ``all one has to do''
is
\\ 
(a) To fix a protocol, \ie a procedure, to transform 
the forces acting
on the system along a well defined {\it fixed once and for all} path
from the initial values to the final values in a fixed time interval
($t=1$ in some units), and 
\\ 
(b) Measure the energy variation $W$ generated by the machines
implementing the protocol. This is a really measurable quantity at
least when $W$ can be interpreted as the work done on
the system, or related to it.
\\
(c) Then average of the exponential of $-\b W$ with respect to a large
number of repetition of the protocol abd apply \equ(e3.2). This can be
useful even, and perhaps mainly, in biological experiments.

\0(ii) Imagine a protocol consisting in lifting a container with a gas
in equilibrium to height $z$: the Hamiltonian changes by $Mgz$, if $M$
is the total mass and $g$ gravity constant. Eq. \equ(e3.2) of course
is correct, being an identity, and gives a free energy variation equal
to $\b M g z$ while normally one would say that the free energy, and
every other thermodynamic quantity, should have remained
unchanged. Whether or not there has been a free energy variation
really depends on what one is interested in studying. Thus if the
interest is in measuring free energy variations in a biology
experiment care has to be given (and is actually given) to make sure
that the protocol followed does not introduce spurious, quite hidden,
forms of work. This makes, once more, clear that the application of a
mathematical identity to real systems requires careful
examination of the conclusions drawn.  \*

The two formulae \equ(e3.2) and \equ(e1.7) bear some similarities but
are, however, quite different:

(1) the $\ig_0^\t \s(S_tx)\, dt$ in \equ(e1.7) is an entropy creation
rather than the energy variation $W$.

(2) the average in \equ(e1.7) is over the SRB distribution of a
    stationary state, in general out of equilibrium, rather than on a
    canonical equilibrium state.

(3) the \equ(e1.7) says that $\media{ e^{-\ig_0^\t
    \e(S_tx)\,dt}}_{SRB}$ is bounded as $\t\to\io$ rather than being
$1$ exactly unlike \equ(e3.2) which holds without corrections,
\cite{Ja97,Ja99}.  
\*

The \equ(e3.2) has proved useful in various equilibrium problems (to
evaluate the free energy variation when an equilibrium state with
Hamiltonian $H_0$ is compared to one with Hamiltonian $H_1$); hence it has
some interest to investigate whether \equ(e1.7) can have some
consequences.

If a system is in a steady state and produces entropy at rate $\s_+$
(\eg a living organism feeding on a background) the fluctuation
relation \equ(e1.6) and its consequence Bonetto's formula, \equ(e1.7),
gives us informations on the fluctuations of entropy production, \ie
of heat produced, and \equ(e1.7) {\it could be useful}, for instance,
to check that all relevant heat transfers have been properly taken
into account.  This suggests that the fluctuation relation for
stationary states could have some applications even in experiments in
biology and be a valuable complement to \equ(e3.2).  \*

\0(3) Finally the identification of entropy creation and pase space
contraction suggests a possible quantitative measure for ``how
irreversible'' is a transformation between two different stationary
states (equilibrium or not). Since physical processes are often
accompanied by volume changes with time $V\to V_t$ it is natural to
allow them and to change the definition of phase space contraction
\equ(e2.2) to, \cite{Ga06},

\kern-5mm
$$\s^\G(\V X)=\s_0(\V X)+\dot U-N\fra{\dot V_t}{V_t}\Eq(e3.3)$$
where $N$ is the number of particles in the volume $V$.

Then one can try to define the ``irreversibility time scale'' $\t(\P)$
for a process $\P$ measuring the time scale over which the process
manifests its irreversibility. Suppose
that in the process $\P$ the parameters controlling the forces
change with time from an initial value $\V F_0$ to a final one $\V
F_\io$; to be definite suppose 
$$\V F(t)=\V F_0+(1-e^{-\g t})(\V F_\io-\V F_0)\Eq(e3.4)$$
The \equ(e3.4) allows us to consider the {\it quasi instantaneous}
changes ($\g\to\io$) as well as the {\it quasi static} ones
($\g\to0$).

Starting the $N$--particles system in the stationary state $\m_0$ with
parameters $\V F_0$ it evolves to $\m_t$ and, eventually, to the
stationary state $\m_\io$ with parameters $\V F_\io$.

Let $\m_{srb,t}=$ be the SRB distribution with parameters $\V F(t)$
``frozen'' at value taken at time $t$. Then if $\s^{srb}_t$ is the
entropy creation rate in the ``frozen'' state $\m_{srb,t}$ an {\it
irreversibility time scale} for $\P$ could be defined as

\kern-8mm
$$\t(\P)^{-1}=\fra1{N^2}\ig_0^\io
\Big(\media{\s^\G_t}_{\m_t}-\media{\s^{srb}_{t}}_{SRB,t}\Big)^2
dt\Eq(e3.5)$$

\kern-6mm
which can be checked to give the ``expected results'' in simple cases
like the Joule expansion, see \cite{Ga06}. Quasi instantaneous
processes $\P$ have a short $\t(\G)$, meaning that irreversibility
becomes noticeable immediately, while quasi static processes have a
long $\t(\P)$ indicating the opposite situation.
\*

\0(4) Although the above analysis is restricted to particle systems it
can be extended to more general systems, in particular to fluids and
turbulence. Not surprisingly as turbulence has been a source of
inspiration for the development of the above ideas, \cite{Ga02,Ga06}.

\*
\0{Aknowledgements: I thank, for comments and suggestions,
  E.G.D Cohen and P. Garrido,}

\revtex

\bibliographystyle{apsrev}

\end{document}